\pdfoutput=1
\documentclass[english]{IEEEtran}
\usepackage[T1]{fontenc}
\usepackage[latin9]{inputenc}
\usepackage{color}
\usepackage{array}
\usepackage{units}
\usepackage{multirow}
\usepackage{amsmath}
\usepackage{amssymb}
\usepackage{graphicx}

\makeatletter

\providecommand{\tabularnewline}{\\}

\usepackage{cite}
\usepackage{fixltx2e}
\usepackage{url}
\usepackage{float}

\makeatother

\usepackage{babel}
\begin{document}

\title{Estimating Cascading Failure Risk with Random Chemistry}

\author{Pooya Rezaei, \emph{Student Member, IEEE}, Paul D. H. Hines, \emph{Senior
Member, IEEE}, Margaret J. Eppstein%
\thanks{This work was supported in part by U.S. Dept.~of Energy award \#DE-OE0000447
and by U.S. National Science Foundation Award \#ECCS-1254549. 

The authors are with the College of Engineering and Mathematical Sciences,
University of Vermont, Burlington, VT (e-mail: \{pooya.rezaei,paul.hines,maggie.eppstein\}@uvm.edu).%
}}
\maketitle
\begin{abstract}
The potential for cascading failure in power systems adds substantially
to overall reliability risk. Monte~Carlo sampling can be used with
a power system model to estimate this impact, but doing so is computationally
expensive. This paper presents a new approach to estimating the risk
of large cascading blackouts triggered by multiple contingencies.
The method uses a search algorithm (Random Chemistry) to identify
blackout-causing contingencies, and then combines the results with
outage probabilities to estimate overall risk. Comparing this approach
with Monte~Carlo sampling for two test cases (the IEEE RTS-96 and
a 2383 bus model of the Polish system) illustrates that the new approach
is at least two orders of magnitude faster than Monte~Carlo, without
introducing measurable bias. Moreover, the approach enables one to
compute the sensitivity of overall blackout risk to individual component-failure
probabilities in the initiating contingency, allowing one to quickly
identify low-cost strategies for reducing risk. By computing the sensitivity
of risk to individual initial outage probabilities for the Polish
system, we found that reducing three line-outage probabilities by
50\% would reduce cascading failure risk by 33\%. Finally, we used
the method to estimate changes in risk as a function of load. Surprisingly,
this calculation illustrates that risk can sometimes decrease as load
increases. \end{abstract}


\section{Introduction}

Large blackouts are low-probability, high-impact events; i.e., they
happen infrequently, but when they do happen, they can lead to catastrophic
social and economic effects, such as the North American blackouts
of Aug.~2003 \cite{:2004} and Sept.~2011 \cite{:2012}. For this
reason cascading failure risk assessment is increasingly required
by reliability standards (e.g., \cite{NERC:2007}) and is a focus
of IEEE Power and Energy Society activities \cite{Vaiman:2012}. 

A primary goal of reliability risk analysis methods is to produce
statistical reliability measures, such as Expected Energy Not Supplied
(EENS), that allow one to compare different versions of a particular
system, or to evaluate changes in a system over time. The science
and practice of Monte~Carlo (MC) methods for power system reliability
studies are well established \cite{Allan:1996}, particularly for
evaluating generation and transmission adequacy. However, for some
types of problems, Monte~Carlo methods can require many samples to
produce high-confidence statistical results, causing the approach
to be computationally prohibitive or even infeasible in some cases
\cite{Borges:2001}. Some studies reduce the computational effort
due to sampling by: 1) variance reduction techniques \cite{Billinton:1997},
2) state-space pruning \cite{Singh:1997}, 3) parallel and distributed
computation \cite{Borges:2001}, 4) pseudo-sequential simulation \cite{DaSilva:2000},
or 5) Artificial Neural Networks \cite{daSilva:2007}. Generally,
the results provide small improvements (typically less than a factor
of 10 speedup). To our knowledge, \cite{daSilva:2007} describes an
approach with the most substantial speedup: a factor of 45 over standard
Monte~Carlo.

Methods for applying sampling techniques to cascading failure risk
estimation are less well established \cite{Vaiman:2012}, for several
reasons. First, the simulation of cascading failures remains a difficult
problem, making it difficult to estimate the eventual impact (typically
measured in the amount of lost load) of a cascade-triggering disturbance.
As a result, standard power system reliability models tend to capture
only one or two outage generations, ignoring the impact of subsequent
outages \cite{Henneaux:2013}. Second, even if cascading sizes can
be accurately estimated, those sizes can be at any scale: from a few
MW to tens of GW. The nature of cascading gives rise to the well-documented
power-law in blackout sizes \cite{Carreras:2000,Carreras:2002}. This
implies that the occurrence probabilities of particular combinations
of outages (and operator errors) that could trigger large cascading
failures are very low, making it necessary to observe several representative
events to obtain an accurate risk estimate \cite{Dobson:2013}. Third,
the size of the search space of all possible $n-k$ contingencies
that might trigger a cascade, where $n$ is the number of components
that might fail and $k$ is the number that did fail, is enormous
and grows exponentially with $n$ and $k$. This further increases
the need for a large number of simulations.

Despite these challenges, there are a number of studies focused on
the modeling of cascading failures (e.g., \cite{Pfitzner:2011,Qi:2013,Rahnamay-Naeini:2014,Henneaux:2013}).
A smaller number of papers have adapted sampling techniques to the
problem of cascading failure risk estimation \cite{Mousavi:2012,Henneaux:2013}.
In doing so, some have used variance reduction techniques to reduce
the computational effort \cite{Kirschen:2004,Chen:2005,Kim:2013,Chen:2013},
which led to a speedup factor of 5-10 in \cite{Kirschen:2004}, and
2-4 in \cite{Chen:2013}. Non-sampling approaches, such as branching
process models \cite{Ren:2008,Dobson:2012}, can provide efficient
estimates of risk, but abstract away some details, such as the ability
to compute the relative contributions of particular outages to \textcolor{black}{overall
risk.}

\textcolor{black}{In our prior work \cite{Eppstein:2012}, we adapted
a search algorithm, known as Random Chemistry (RC), to the problem
of finding large collections of $n-k$ contingencies that lead to
cascading failure. However, this initial work did not explain how
the approach could be used for risk estimation. In this paper, we
derive a method to use the Random Chemistry algorithm to estimate
blackout risk due to cascading failure, given a few simplifying assumptions,
and compare the computational efficiency of our approach with that
of Monte~Carlo sampling. This paper builds on preliminary results
reported in \cite{Rezaei:2014} and \cite{Rezaei:2014aa}, extending
our method to compute the relative impacts of individual components
on overall risk and showing how increasing load and generator dispatch
method impact risk.}

\textit{\emph{The remainder of the paper is organized as follows.
Section \ref{sec:Two-approaches} defines cascading failure risk and
introduces two general approaches to estimate risk. Section \ref{sec:RC-Method}
describes our method for using Random Chemistry for cascading failure
risk estimation, and section \ref{sec:Results} presents simulation
results. Finally, section \ref{sec:Conclusions} provides our conclusions. }}

\section{Two approaches to risk estimation\label{sec:Two-approaches}}

A standard measure of risk due to a random disturbance is the product
of event probability and its severity (or cost) \cite{Vaiman:2012}.
If $S(\mathcal{C},x)$ is a measure of the severity of the events
proceeding from an arbitrary disturbance $\mathcal{C}$ to a system
at a particular state $x$, then the risk due to $\mathcal{C}$ is:

\begin{equation}
R(\mathcal{C},x)=\Pr(\mathcal{C})S(\mathcal{C},x)
\end{equation}
If we denote the set of all possible disturbances by $\Omega$, then
the system risk is: 

\begin{equation}
R(x)=\sum_{\forall c\in\Omega}R(c,x)=\sum_{\forall c\in\Omega}\Pr(c)S(c,x)\label{eq:true_R}
\end{equation}
In this paper, $\mathcal{C}$ denotes a random disturbance, while
$c$ (the lower case) denotes an individual disturbance. Therefore,
one can think of $S(\mathcal{C},x)$ as a random variable that maps
a random event $\mathcal{C}$ to its severity, and interpret $R(x)$
as $E[S(\mathcal{C},x)]$, i.e., the expected value of $S(\mathcal{C},x)$.

Since it may not be feasible to compute the severity of all disturbances
for a system, one approach for estimating system risk is to estimate
$R(x)$ from a smaller sample of all possible disturbances. One option
is Monte~Carlo sampling, in which one draws randomly from $\Omega$
based on the probability function $\Pr(\mathcal{C})$. The average
severity then converges to the true risk, given a sufficiently large
set of samples: 
\begin{equation}
\hat{R}_{MC}(x)=\frac{1}{|\Omega_{a}|}\sum_{c\in\Omega_{a}}S(c,x)\label{eq:R_MC}
\end{equation}
where $\hat{R}$ is an estimated value of $R$ in this and subsequent
notations, $\Omega_{a}$ is a set that results from sampling randomly
from $\Omega$, and $\left|\Omega_{a}\right|$ is the number of elements
in $\Omega_{a}$. Note that $\Omega_{a}$ may include duplicates and
the empty set, depending on the sampling approach. 

A key problem with this approach is that there are many events for
which $S(c,x)=0$, which means that Monte~Carlo will spend a lot
of time sampling from events that have no impact on $R(x)$. Alternatively,
if one could sample exclusively from the subspace $\Omega_{B}$ such
that $S(c,x)>0,\forall c\in\Omega_{B}$, and we somehow knew the size
of this subset, then risk could be more efficiently estimated by computing:

\begin{equation}
\hat{R}(x)=\frac{|\Omega_{B}|}{|\Omega_{b}|}\sum_{c\in\Omega_{b}}\Pr(c)S(c,x)\label{eq:R_RC}
\end{equation}
where $\Omega_{b}$ is a representative subset sampled from $\Omega_{B}$
, i.e., $\Omega_{b}\subseteq\Omega_{B}$. Clearly, (\ref{eq:R_RC})
converges to (\ref{eq:true_R}) as $\Omega_{b}\rightarrow\Omega_{B}$.
In order to avoid a biased outcome when $|\Omega_{b}|\ll|\Omega_{B}|$,
$\Omega_{b}$ needs to be an unbiased sample, such that it provides
a representative sample of $\Omega_{B}$ across a range of event sizes
and probabilities. The idea in (\ref{eq:R_RC}) is somewhat similar
to importance sampling \cite{Cochran:1963}.

In this paper, we develop a method for estimating cascading blackout
risk in power systems based on (\ref{eq:R_RC}), and compare the convergence
speed with Monte~Carlo sampling in (\ref{eq:R_MC}).

\section{Method\label{sec:RC-Method}}

The specific focus of this paper is estimating risk due to cascading
failures triggered by exogenously-caused branch (transmission line
or transformer) outage contingencies. In this case, each contingency
$c\in\Omega$ is a subset of the branches in the grid ($c\subseteq\{1,...,n\}$),
and $\Omega$ is the set of all possible branch outage combinations.
While there are other types of contingencies that might trigger cascades,
branch outages provide a useful starting point. Adapting the method
to incorporate other type of contingencies (e.g., generator outages)
is trivial. In this paper, we assume that the exogenously-caused branch
outages are statistically independent, such that for two single contingencies
$c_{i}$ and $c_{j}$, $\Pr(c_{i}\cap c_{j})=\Pr(c_{i})\Pr(c_{j})$.
While there are methods in the literature for modeling common mode
outages (e.g., \cite{Grigg:1999}) and other types of correlations
(e.g., \cite{Nedic:2006}), the risk from these correlations is not
captured in the results presented here. Verifying the efficiency of
the Random Chemistry risk estimation method for the case of correlated
outages and other types of contingencies remains for future work.

In this paper, $S(c,x)$ represents the output of a simulator that
gives blackout sizes in MW, resulting from any initiating contingency
$c$ applied to the system with state $x$ such that $0\leq S(c,x)\leq S_{0}$,
where $S_{0}$ is the total load in the system. In general, Monte~Carlo
sampling can quickly estimate $R(x)$ for blackouts of small sizes.
However, because of the heavy-tailed statistics of cascading failures,
estimating the risk due to larger blackouts is more difficult. Here,
we focus on computing the risk resulting from blackouts larger than
some fraction $0<\alpha<1$ of $S_{0}$. Specifically, we study the
risk of blackouts 5\% and larger, i.e., for $S(\mathcal{C},x)\geq0.05S_{0}$.
In order to do so, we replace $S(\mathcal{C},x)$ in (\ref{eq:R_MC})
with $S_{\alpha}(\mathcal{C},x)$, where: 

\[
S_{\alpha}(\mathcal{C},x)=\left\{ \begin{array}{ccc}
S(\mathcal{C},x) &  & S(\mathcal{C},x)\geq\alpha S_{0}\\
0 &  & \textrm{otherwise}
\end{array}\right.
\]

\subsection{Risk from minimal contingencies\label{sub:Random-Chemistry-Risk}}

A unique aspect of the Random Chemistry algorithm (see Appendix I
and \cite{Eppstein:2012}) is that it is designed to find ``minimal''
blackout-causing contingencies. A minimal blackout-causing contingency
is a set of outages that result in a large blackout (in this paper
we define large as $\geq5\%$), but would not result in a large blackout
if any one of the outages in the multiple-contingency had not occurred.
For brevity we refer to disturbances of this sort as malignancies.
\textcolor{black}{Formally, a set of outages $d$ is a malignancy
if $S(d,x)\geq0.05S_{0}$ and $\nexists c'\in\Omega:c'\subset d\wedge S(c',x)\geq0.05S_{0}$.
In order to use the data from Random Chemistry, we need to be able
to estimate the risk from all $n-k$ contingencies, including the
non-minimal ones, from data about malignancies. }To do so, we make
the following assumption:

\textbf{Assumption 1}: \emph{Given a malignancy }\textcolor{black}{$d$}\emph{,
which triggers a blackout of size $S(d,x)$, the blackout sizes triggered
by all supersets of }\textcolor{black}{$d$}\emph{ }\textit{can be
estimated by the size of blackout triggered by }\textcolor{black}{$d$}\emph{,
i.e., $\forall c''\supset d\rightarrow S(c'',x)=S(d,x)$.}

Clearly, Assumption 1 does not hold for every case. There are some
superset contingencies that trigger smaller blackouts and some supersets
that cause larger blackouts. In Appendix II, we evaluate Assumption
1, comparing the actual blackout sizes of all $n-3$ supersets with
the values that are estimated by minimal $n-2$ subsets for the Polish
system (described in section \ref{sec:Results}). We find that Assumption
1 causes the $n-3$ risk estimate to be 2.5\% higher than the actual
value for $n-3$ risk. Given that the $n-3$ risk is 24.0\% of the
total risk, the total impact of Assumption 1 on the overall risk estimate
is to increase it by only 0.6\%. 

As one adds more components to a minimal malignancy, making larger
supersets, the blackout size difference between the two events becomes
larger, but because the probability of high order events ($n-k$ for
$k\geq4$) is orders of magnitude lower than probability of its malignancy
subset, the overall impact becomes much smaller, and at some point
negligible for high order events.

This section shows, by a small example, how to use Assumption 1 to
estimate risk efficiently using the minimal contingency data that
Random Chemistry provides. 

Let us assume that in a small system with 5 transmission lines, Random
Chemistry has found an $n-3$ malignancy, $d_{1}=\{1,2,3\}$. (By
definition, this means that none of the $n-2$ subset contingencies,
$c'_{1}=\{1,2\}$, $c'_{2}=\{1,3\}$ and $c'_{3}=\{2,3\}$, cause
a 5\% or larger blackout.) Using our previous assumption that line
failures are independent, the exact risk associated with $d_{1}$
and all its supersets is:

\setlength\abovedisplayskip{0pt} 

\begin{align}
R(\hat{d_{1}},x)= & S(d_{1},x)p_{1}p_{2}p_{3}(1-p_{4})(1-p_{5})+\nonumber \\
 & S(c''_{1},x)p_{1}p_{2}p_{3}p_{4}(1-p_{5})+\nonumber \\
 & S(c''_{2},x)p_{1}p_{2}p_{3}(1-p_{4})p_{5}+\nonumber \\
 & S(c''_{3},x)p_{1}p_{2}p_{3}p_{4}p_{5}
\end{align}
where $\hat{d_{1}}$ denotes $d_{1}$ together with all its supersets:
$c''_{1}=\{1,2,3,4\}$, $c''_{2}=\{1,2,3,5\}$ , $c''_{3}=\{1,2,3,4,5\}$,
and $p_{i}$ is the line-failure probability for line $i$. Given
Assumption 1, $S(d_{1},x)=S(c''_{i},x),\forall i\in\{1,2,3\}$, which
means that:

\begin{multline}
R(\hat{d_{1}},x)=S(d_{1},x)p_{1}p_{2}p_{3}[(1-p_{4})(1-p_{5})+\\
p_{4}(1-p_{5})+(1-p_{4})p_{5}+p_{4}p_{5}]
\end{multline}
Expanding the term in the bracket shows that it equals 1, which means
that:

\begin{equation}
R(\hat{d_{1}},x)=S(d_{1},x)p_{1}p_{2}p_{3}=S(d_{1},x)\Pr(\hat{d_{1}})\label{eq:R(c1,x)}
\end{equation}
where $\Pr(\hat{d_{1}})$ is the probability of the malignancy $d_{1}$
and all of its supersets, i.e., any event that includes at least the
branches in $d_{1}$. A similar formulation holds for supersets of
any malignancy. Therefore, Assumption 1 enables us to estimate risk
of a malignancy and its supersets solely using the information provided
by the malignancy itself. In general, for each malignancy $d$, we
have: 

\begin{equation}
R(\hat{d},x)=S(d,x)\Pr(\hat{d})=S(d,x)\left(\prod_{i\in d}p_{i}\right)\label{eq:R(d,x)_general}
\end{equation}

\subsection{General Random Chemistry risk estimation}

Our method separately estimates the risk due to $n-2$ and $n-3$
malignancies. We use the symbol $\Omega_{m,k}\subset\Omega$ to be
the set of all $n-k$ malignancies. After $i$ runs, Random Chemistry
will have found a certain number of events from each $\Omega_{m,k}$
set. Let $i_{k}$ be the number of times that Random Chemistry has
found a not-necessarily-unique member of $\Omega_{m,k}$ and $\Omega_{RC,k}$
be the set of unique members found. By tracking the rate at which
unique members of $\Omega_{m,k}$ are found, if $i_{k}$ is sufficiently
large, we can estimate the size of $\Omega_{m,k}$ (see section \ref{sub:Estimating-the-Size}).
Let $\hat{m_{k}}$ be this estimate. Given this notation, and the
derivation in (\ref{eq:R(c1,x)}), we can simultaneously capture the
$n-k$ risk from a set of malignancies and all of their supersets
based on (\ref{eq:R_RC}) and (\ref{eq:R(d,x)_general}), as follows:

\begin{equation}
\hat{R}_{RC,k}(x)=\frac{\hat{m}_{k}}{|\Omega_{RC,k}|}\sum_{d\in\Omega_{RC,k}}S(d,x)\left(\prod_{i\in d}p_{i}\right)\label{eq:risk-RC_k}
\end{equation}
where $\hat{R}_{RC,k}(x)$ is the estimated $n-k$ blackout risk by
Random Chemistry. The results in section \ref{sec:Results} indicate
that the sum of these values over $k$ gives a reasonable estimate
of total blackout risk:

\begin{equation}
\hat{R}_{RC}(x)=\sum_{k=2}^{k_{\max}}\hat{R}_{RC,k}(x)\label{eq:risk-RC}
\end{equation}
We have set $k_{\textrm{max}}=5$ while running Random Chemistry previously,
since larger multiple contingencies are highly improbable, and (at
least for the case of uncorrelated outages) have a minuscule contribution
to the total risk. However, the size of the $n-5$ and $n-4$ sets
are so large that for real-world systems, it is not computationally
practical to estimate $\hat{m}_{5}$ and $\hat{m}_{4}$. Thus we generally
use $k_{\max}=3$ for risk calculations. Since this captures the risk
from an increasingly large collection of $n-\{2,3\}$ malignancies
and \emph{all of their supersets}, the results appear to be highly
representative of the true system risk (see section~\ref{sec:Results}). 

Note that there is a caveat in considering the malignancy supersets
as in (\ref{eq:risk-RC_k}). Every contingency set causing a large
blackout contains a malignancy, but this malignancy is not necessarily
unique. The contingencies that contain more than one unique malignancy
are considered more than once in (\ref{eq:risk-RC_k}). For example,
if \{1,2\} and \{1,3\} were both malignancies in a system, the superset
\{1,2,3\} would be taken into account two times in (\ref{eq:risk-RC_k});
once in $S(\{1,2\},x)p_{1}p_{2}$ and once in $S(\{1,3\},x)p_{1}p_{3}$.
However, the number of these kinds of supersets is normally much lower
than the total number of supersets of any order, so the overall error
that they cause is negligible. In this paper, we are mainly studying
the $n-2$ and $n-3$ risk, so we counted the exact number of $n-3$
supersets that have been counted more than once in the Polish system.
Our inspection indicated that out of the total of 1,562,760 $n-3$
supersets of $n-2$ malignancies, 21,771 supersets (1.4\%) included
more than one malignancy subset, and thus were counted more than once.
Note that the exact number of malignancies that will be counted more
than once depends on the particular system. It is also noteworthy
that each of the higher order supersets may include more malignancies
than each of the lower order ones, however, as mentioned before, the
higher order contingencies have orders of magnitude lower probabilities
and thus have an infinitesimal impact on the overall risk. 

It is important to note that we find, in section \ref{sec:Results},
that Assumption 1 and the above caveat do not measurably change the
risk estimation results relative to the Monte~Carlo estimate, which
computes risk based on the exact probability and blackout sizes of
all contingencies.

\subsection{Estimating the Size of Each $n-k$ Collection\label{sub:Estimating-the-Size}}

\begin{figure}
\begin{centering}
\includegraphics[width=1\columnwidth]{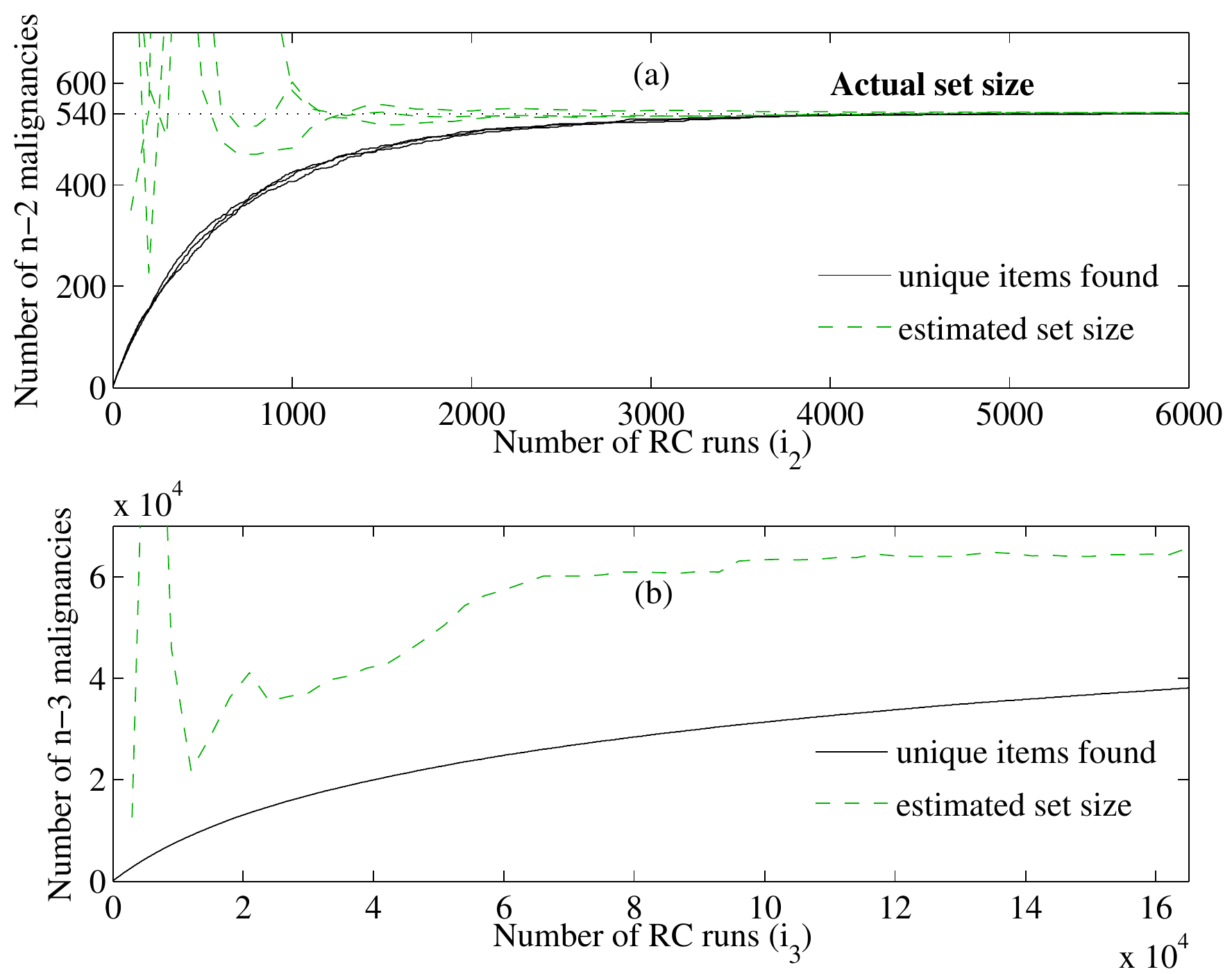}
\par\end{centering}

\protect\caption{The number of unique (a) $n-2$, and (b) $n-3$ malignancies found
($\Omega_{RC,k}$,solid line) and the estimate of the set size ($\hat{m_{k}}$,
dashed line) vs.~the number of Random Chemistry runs that led to
an $n-2$ ($i_{2}$) and $n-3$ ($i_{3}$) malignancy, respectively,
in the Polish system.\label{fig:simulate_rc}}
\end{figure}

Initially, Random Chemistry primarily finds malignancies that are
unique (not found previously). However, as the number of identified
malignancies increases, the algorithm finds duplicates more frequently.
By measuring the rate at which the algorithm finds new unique contingencies
of a given size $k$, one can estimate the total number of minimal
dangerous $n-k$ contingencies that cause blackouts. There are a variety
of ways to estimate the set-size $|\Omega_{m,k}|$ as Random Chemistry
progresses, such that:$\underset{i_{k}\rightarrow\infty}{\lim}\hat{m}_{k}(i_{k})=\left|\Omega_{m,k}\right|$.
In \cite{Eppstein:2012}, we suggested an approach based on measuring
the rate at which the algorithm is finding new malignancies. In this
paper, we use another approach based on the number of unique malignancies
found vs.~the number of Random Chemistry runs. To illustrate, consider
a jar of $N$ balls numbered $1,...,N$. If balls are removed from
the jar one at a time with the same probability (and then replaced),
one can mathematically show that the expected number of unique balls
drawn after $i$ draws ($N_{i}$) follows:
\begin{equation}
N_{i}=N(1-e^{ri})
\end{equation}
where $r=\ln\left(1-\frac{1}{N}\right)$. Based on this idea, we suggest
an alternative exponential model, which was empirically found to be
more accurate for our application. We found that the number of unique
malignancies found by Random Chemistry is best represented by the
Cumulative Distribution Function (CDF) of the exponential Weibull
distribution. Specifically, the number of unique $n-k$ malignancies
in the set $\Omega_{RC,k}$ after $i_{k}$ runs follows:

\begin{equation}
|\Omega_{RC,k}(i_{k})|\sim\hat{m_{k}}\left(1-e^{-(\frac{i_{k}}{\lambda})^{\mu}}\right)^{\nu}\label{eq:mk-1}
\end{equation}
The parameters $\lambda$, $\mu$, $\nu$ and $\hat{m}_{k}$ (the
estimate of interest) can be found by nonlinear least-squares fitting
to a sequence of data for $i_{k}$ and $|\Omega_{RC,k}(i_{k})|$.
Fig.~\ref{fig:simulate_rc} shows the results of this method, where
the the sizes of $n-2$ and $n-3$ malignancy sets are estimated for
the Polish system. The results shown for $n-2$ are for three separate
runs of Random Chemistry searching for malignancies in the Polish
system (Fig.~\ref{fig:simulate_rc}a). All three estimates converged
to $\hat{m_{k}}=540$, which is the true number of $n-2$ malignancies,
as verified by exhaustive search. Fig.~\ref{fig:simulate_rc}b shows
that the estimate of $\hat{m_{3}}$ is converging to about $6.4\times10^{4}$.

\subsection{Sensitivity of risk to component-failure probability\label{sub:Finding-the-contributions-1}}

As previously suggested, it is possible to use the Random Chemistry
risk estimation approach in (\ref{eq:risk-RC_k}) to find the sensitivity
of the overall risk to individual component-failure probability in
the initiating contingency. To do so, we compute the partial derivative
of $\hat{R}_{RC,k}(x)$ with respect to each branch failure probability
$p_{i}$ and sum the result over $k\in\{2,3\}$ for all branches,
as follows:

\begin{eqnarray} 
\frac{\partial\hat{R}_{RC,k}(x)}{\partial p_{i}} 
& = 
& \frac{\hat{m}_{k}}{|\Omega_{RC,k}|}\sum_{d\in\Omega_{RC,k} }S(d,x)\left(\prod_{\substack{j\in d \\ j\neq i}}p_{j}\right)
\end{eqnarray}

\begin{equation}
\frac{\partial\hat{R}_{RC}(x)}{\partial p_{i}}=\sum_{k=2}^{k_{\max}}\frac{\partial\hat{R}_{RC,k}(x)}{\partial p_{i}}\label{eq:sensitivity}
\end{equation}

This type of sensitivity analysis is particularly useful in finding
critical components in a transmission system (see section~\ref{sub:Finding-the-contributions}).
Note that (\ref{eq:sensitivity}) provides the sensitivity of risk
to individual outage probability in the initiating events, and to
keep this paper focused, we do not study the failure probabilities
when a cascade propagates. In future work, we will identify the lines
that fail frequently during the course of cascades, and study the
impact on risk of upgrading these lines.

\section{Results\label{sec:Results}}

We used two test systems to evaluate the proposed risk estimation
method. The first is the 73-bus RTS-96, which has 120 branches and
8550 MW of total load \cite{Grigg:1999}. The second test system is
a model of the 2004 winter peak Polish power system, which is available
with MATPOWER \cite{Zimmerman:2011}. This test system has 2896 branches
(transmission lines and transformers), 2383 buses, and 24.6 GW of
total load. For the Polish case, some of the transmission lines were
overloaded in the original system, so we increased line flow limits
to be the larger of the existing limit and 1.05 times the pre-contingency
line flows for each line, after increasing all loads by 10\%. We prepared
the pre-contingency test cases for both systems using a preventive
security-constrained dc optimal power flow (SCDCOPF) to make both
cases $n-1$ secure with respect to branch outages (see~\cite{Rezaei:2014aa}
for details). The $n-1$ security constraints in the SCDCOPF model
ensure that all flows remain less than or equal to the long-term emergency
(LTE) rating after a single branch outage. The LTE rating is assumed
to be 110\% of the normal flow limit for all lines. The generation
cost data, which is needed for the OPF, is not readily available for
RTS-96. In this paper, we took the cost data from \cite{EIA} based
on the generation types in \cite{Grigg:1999}.

In both networks, we computed line failure probabilities from transmission
line failure rates ($\lambda$, outages/year), and assumed that each
failure lasts for 1 hour on average (that the mean time to repair
is 1 hour). Thus, the probability of line $i$ being in a non-working
state during a particular sample is $p_{i}=\nicefrac{\lambda_{i}}{8760}$,
where 8760 is the number of hours in a year. Note that this approximation
is valid only for the case where $\nicefrac{\lambda_{i}}{8760}\ll1$,
which is true, since all of the $\lambda$ in the test systems are
approximately 1. As the failure rate data were not available for the
Polish system, we randomly assigned failure rates to transmission
lines, such that the mean and variance of $\lambda$'s were equal
to those in the RTS-96.

\subsection{Simulator}

To test our method, we used an updated version of the cascading failure
simulator used in \cite{Eppstein:2012}. In this simulator (known
as DCSIMSEP) line flows are computed using a dc power flow calculation.
When outages result in long-term emergency (LTE) rating violations
(assumed to be 110\% of the normal line flow limit), transmission
lines trip in an amount of time that is proportional to the magnitude
of the overload, such that more extreme overloads result in faster
outages. The time delay is set such that if the line flow is 50\%
above the normal limit (that is the short-term emergency rating, STE),
the line will trip in 5 seconds. When line outages result in the separation
of a network into multiple connected components (islands), a combination
of fast generator ramping, load shedding and generator tripping are
used to restore a balance between supply and demand in each island.
After each simulation, DCSIMSEP reports the number of line outages
and the amount of load shedding that occurred. Illustrative video
and source code for the simulator are available at \cite{DCSIMSEP}.

DCSIMSEP does not describe several known mechanisms of cascading failure,
such as voltage collapse, dynamic instability or hidden failures in
protection systems. However, the risk analysis process proposed here
can be used in combination with any simulator that reports the blackout
size, given an initiating contingency. Future work will test the impact
of these additional mechanisms on the efficiency of the proposed approach.

\subsection{Comparing Random Chemistry approach to Monte~Carlo\label{sub:Comparing-RC-approach}}

We applied the proposed method to our two test cases, and estimated
large blackout risk as a function of total computational effort. These
results were compared to results from a standard Monte~Carlo simulation.
The number of calls to DCSIMSEP is used as our measure of computational
effort, because cascading failure simulation takes up the vast majority
of risk estimation run-time. Actual run-time is not used explicitly
since it depends on many factors, including the extent to which code
is optimized. Each cascading failure simulation (by a call to DCSIMSEP)
in the risk estimation process takes 0.3 to 0.5 seconds on average
for the Polish system%
\footnote{On a 2.66 GHz Intel Core i7 MacBook Pro with 8 GB of memory%
}, depending on the number of subsequent cascades. In our Monte~Carlo
implementation, a contingency is simulated only if the number of elements
in the sampled contingency is 2 or larger, since the system was initially
$n-1$ secure. While each Monte~Carlo run needs either zero or one
simulation, each successful Random Chemistry run requires approximately
47 simulations (when averaged over both successful and unsuccessful
runs) on the Polish system. 

\begin{figure*}
\begin{centering}
\includegraphics[width=1\textwidth]{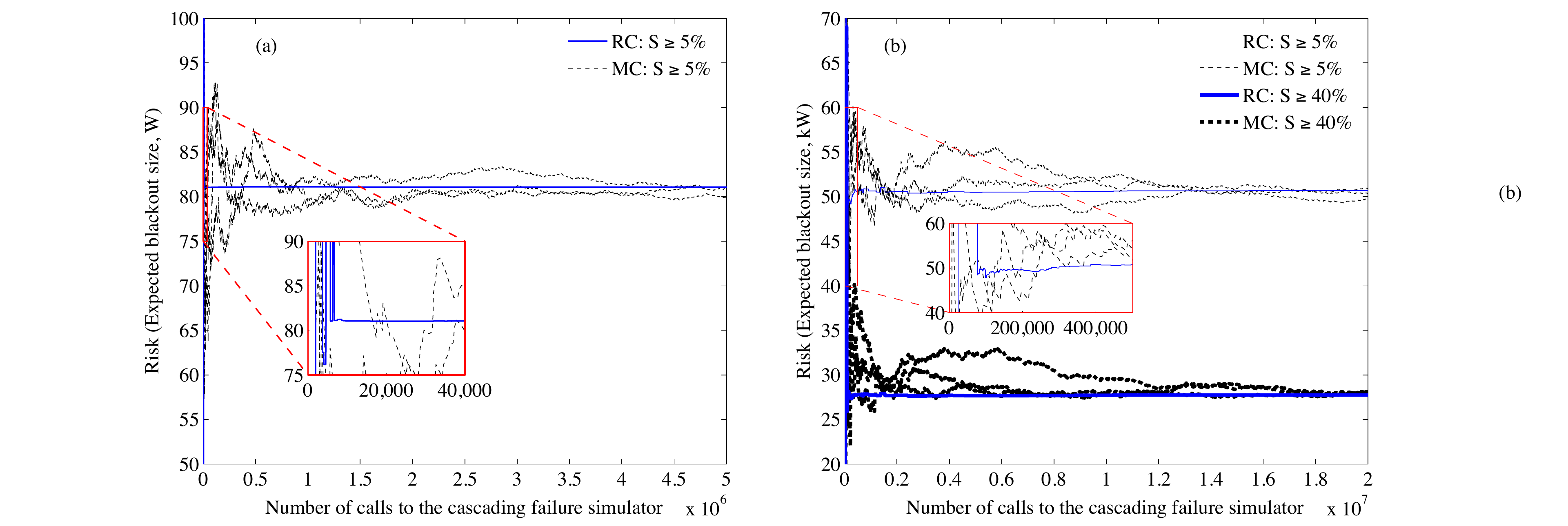}
\par\end{centering}

\protect\caption{\label{fig:mc_and_rc}Cascading failure risk estimates using the Random
Chemistry and Monte~Carlo methods for the (a) RTS-96 and (b) Polish
system, where $S$ is the blackout size. For each test case, three
representative Monte~Carlo runs and one Random Chemistry run (since
runs for the latter are very consistent) are shown. }
\end{figure*}

Figs.~\ref{fig:mc_and_rc}a and \ref{fig:mc_and_rc}b show the Monte~Carlo
and Random Chemistry risk convergence for the RTS-96 and the Polish
systems, respectively. More specifically, Fig.~\ref{fig:mc_and_rc}a
shows the convergence of $E[S_{0.05}(\mathcal{C},x)]$ for the RTS-96,
while Fig.~\ref{fig:mc_and_rc}b shows the convergence of $E[S_{0.05}(\mathcal{C},x)]$
and $E[S_{0.4}(\mathcal{C},x)]$ for the Polish system. In the case
of RTS-96, the largest blackout caused by $n-2$ and $n-3$ malignancies
was less than $20\%$ of total load, therefore we only explored $E[S_{0.05}(\mathcal{C},x)]$. 

Fig. \ref{fig:mc_and_rc} shows that the Random Chemistry approach
is more than two orders of magnitude faster than Monte~Carlo for
both test systems. We also observe that Monte~Carlo and Random Chemistry
are converging to similar values, providing evidence that Assumption
1 does not measurably bias the results. 

In addition, Table \ref{tab:mc_rc} compares the convergence numerically
for $S\geq5\%$ in the Polish system. We computed the mean, variance
and 95\% confidence intervals (the range between $2.5^{\textrm{th}}$
and $97.5^{\textrm{th}}$ percentiles) for both Random Chemistry and
Monte~Carlo by bootstrapping, where random samples were taken with
replacement from the original data. Results are for 1000 resampling
trials, where for each trial, the number of samples were chosen for
both Random Chemistry and Monte~Carlo such that they required the
same number of calls to the cascading failure simulator. 

\begin{table}[h]
\protect\caption{\label{tab:mc_rc}Comparison of Monte~Carlo (MC) and Random Chemistry
(RC) risk (in kW) after $10^{6}$ and $2\times10^{7}$ calls to the
simulator for 1000 bootstrapping trials}

\centering{}%
\begin{tabular}{>{\centering}p{0.25\columnwidth}>{\centering}p{0.1\columnwidth}>{\centering}p{0.1\columnwidth}>{\centering}p{0.1\columnwidth}>{\centering}p{0.1\columnwidth}}
\cline{2-5} 
 & \multicolumn{2}{c}{After $10^{6}$ calls} & \multicolumn{2}{c}{After $2\times10^{7}$ calls}\tabularnewline
\cline{2-5} 
 & MC & RC & MC & RC\tabularnewline
\hline 
Mean & 53.03 & 50.25 & 50.32 & 50.48\tabularnewline
\hline 
Variance & 21.87 & 0.007 & 0.95 & 0.000045\tabularnewline
\hline 
95\% Confidence Interval & 44.54-62.84 & 50.07-50.36 & 48.34-52.29 & 50.46-50.49\tabularnewline
\hline 
\end{tabular}
\end{table}

Table \ref{tab:mc_rc} shows that after $10^{6}$ calls to the simulator,
the confidence interval for Monte~Carlo is much wider than for Random
Chemistry. Note that the very minor drift in the Random Chemistry
confidence interval from $10^{6}$ to $2\times10^{7}$ calls to the
simulator is due to the improved estimate of $\hat{m_{3}}$ (Fig.~\ref{fig:simulate_rc}b).

\subsection{Investigating the components of large blackout risk }

There are a number of advantages that arise from computing risk by
sampling methods such as Random Chemistry or Monte~Carlo. With both
methods, one can disaggregate \textcolor{black}{$\hat{R}(x)$ and
compute the risk that comes from blackouts of different sizes, e.g.,
compute the risk due to small, medium, and large blackouts, or the
risk from $n-2$ and $n-3$ contingencies, although it takes much
longer for Monte~Carlo to find a sufficient number of outages to
do this type of disaggregation. Sections \ref{sub:Risk-from-BOs}
and \ref{sub:Risk-from-n=0022122} illustrate this calculation, and
section \ref{sub:Estimating-risk-vs-load-level} shows the disaggregated
(in blackout size) risk vs.~load level. }

\textcolor{black}{A second advantage is the ability to find the sensitivity
of $\hat{R}(x)$ to the failure-rates of particular transmission lines.
}D\textcolor{black}{isaggregat}ing $\hat{R}(x)$ and finding the most
important components that impact $\hat{R}(x)$ could have tremendous
value to system operators. For example, it may allow an operator to
identify particular actions (such as hardening particularly vulnerable
transmission lines, or reducing the load on some lines) that can substantially
reduce large blackout risk.

\subsubsection{Risk from blackouts of different sizes\label{sub:Risk-from-BOs}}

To illustrate the separation of risk for different blackout sizes,
Fig. \ref{fig:mc_and_rc}b shows the risk for $E[S_{0.4}(\mathcal{C},x)]$,
i.e. for blackouts 40\% or larger. Disaggregating risk by blackout
size is particularly useful if one wants to know the impact of potential
system changes, such as the addition of a new transmission line, or
changes in load level. Understanding how these changes impact the
risk of blackouts of different sizes should allow for more informed
decision making. We study the problem of how risk changes with load
in section \ref{sub:Estimating-risk-vs-load-level}.

\subsubsection{Risk from $n-2$ and $n-3$ contingencies\label{sub:Risk-from-n=0022122}}

It is noteworthy that the majority of the risk shown in Fig. \ref{fig:mc_and_rc}
comes from $n-2$ malignancies. Because of the smaller number of $n-2$
malignancies relative to $n-3$ malignancies, the Random Chemistry
$n-2$ risk estimate converges much faster than the $n-3$ risk estimate.
However, well after the $n-2$ risk has converged, some small fluctuations
in total risk remain, due to changes in the minimal $n-3$ set size
estimation (see Fig. \ref{fig:simulate_rc}b). 

In order to further understand the impact of $n-3$ contingencies
on blackout risk, we further explored the details of contingency sets
found for the Polish system. The $n-3$ risk is composed of two portions;
a small part of the risk is due to $n-3$ malignancies (herein referred
to as minimal $n-3$ risk) with the remainder coming from non-minimal
$n-3$ contingencies that are supersets of $n-2$ malignancies. To
find the latter portion, we post-processed the data to compute the
probabilities of all $n-3$ supersets of $n-2$ malignancies, using
the associated $n-2$ blackout size based on Assumption 1, to avoid
simulating all the supersets. The results were similar (see Table
\ref{tab:n_2_3}) to what Monte~Carlo finds when computing risk by
sampling from both minimal and non-minimal contingencies. 

\begin{table}[b]
\begin{centering}
\protect\caption{\label{tab:n_2_3}Estimated \textit{$n-2$} and \textit{$n-3$} risk
in the Polish system.}

\par\end{centering}

\begin{centering}
\begin{tabular}{|c|>{\centering}p{0.13\columnwidth}|>{\centering}p{0.13\columnwidth}cc|}
\cline{2-5} 
\multicolumn{1}{c|}{} & \multirow{2}{0.13\columnwidth}{\centering{}MC risk (kW)} & \multirow{2}{0.13\columnwidth}{\centering{}RC risk (kW)} & \multicolumn{2}{c|}{Percent of RC risk due to}\tabularnewline
\cline{4-5} 
\multicolumn{1}{c|}{} &  &  & $n-2$ minimal & $n-3$ minimal\tabularnewline
\hline 
$n-2$  & 37.1 & 36.4 & 100\% & NA\tabularnewline
\hline 
$n-3$ & 11.5 & 12.1 & 96.7\% & 3.3\%\tabularnewline
\hline 
$\textrm{Total}^{*}$ & 50.3 & 50.5 & 99.2\% & 0.8\%\tabularnewline
\hline 
\end{tabular}
\par\end{centering}

$^{*}$Note that Total includes risk from non-minimal $n-4$'s, $n-5$'s,
etc.
\end{table}

Table \ref{tab:n_2_3} shows the $n-2$ and $n-3$ estimates of risk
by both Monte~Carlo and Random Chemistry methods for the Polish system.
Although the Monte~Carlo risk estimates are similar to those from
Random Chemistry, Monte~Carlo identifies far fewer minimal $n-3$
than non-minimal $n-3$ contingencies, thus providing much less information
about the minimal $n-3$ risk. Specifically, Random Chemistry found
540 $n-2$ and 38,212 $n-3$ malignancies, whereas Monte~Carlo found
only 440 $n-2$, 41 minimal $n-3$, and 830 non-minimal $n-3$ contingencies.
Note that, as Table \ref{tab:n_2_3} shows, 96.7\% of the Random Chemistry
$n-3$ risk estimate (11.7 kW) was actually attributable to $n-2$
malignancies, with only 3.3\% (0.4 kW) due to $n-3$ malignancies.

\subsubsection{Risk Sensitivity\label{sub:Finding-the-contributions}}

\begin{figure}
\begin{centering}
\includegraphics[width=1\columnwidth]{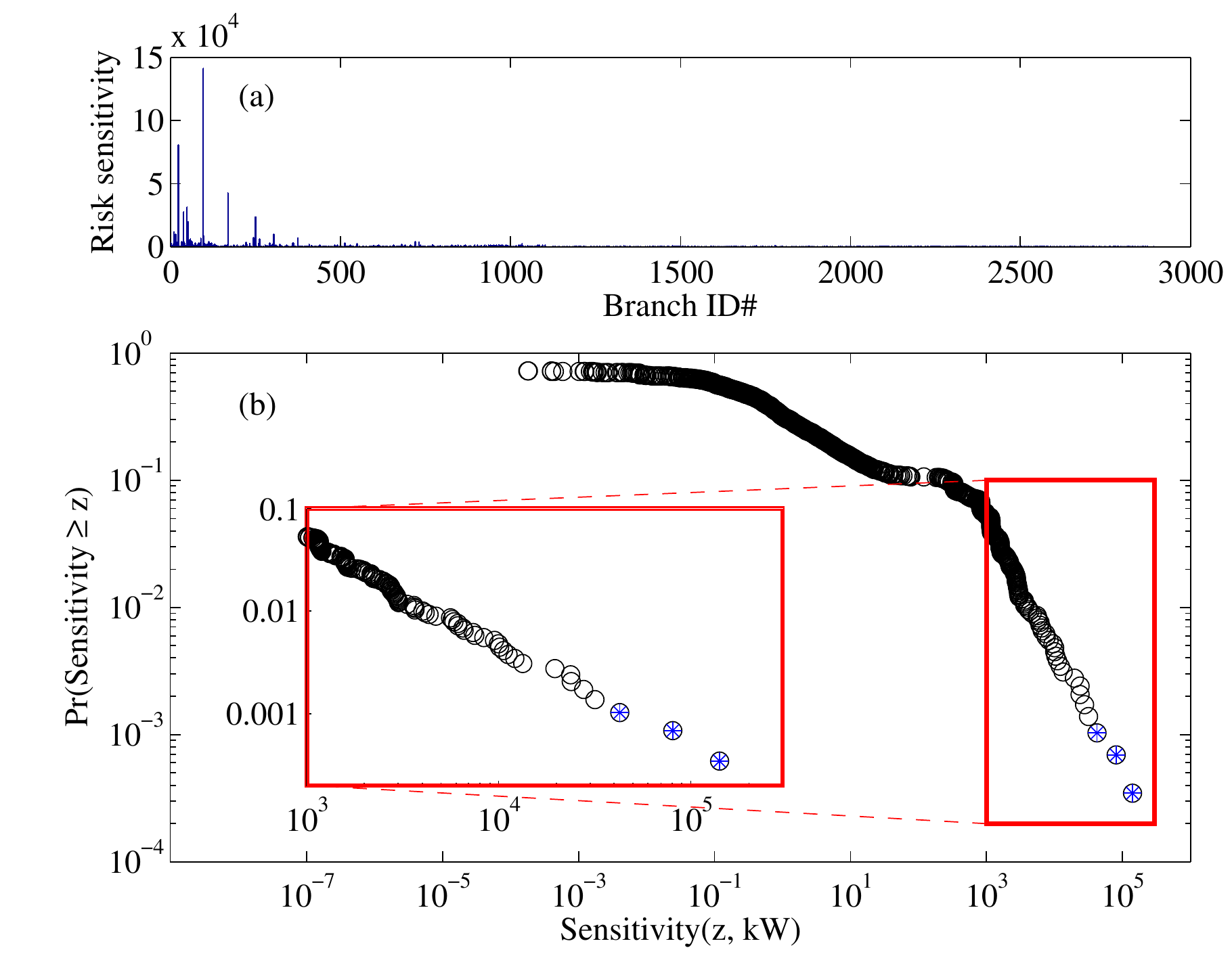}
\par\end{centering}

\protect\caption{\label{fig:Risk-sensitivity}(a) Sensitivity of risk to branch failure
probability for all 2896 branches in the Polish system, (b) Complementary
Cumulative Distribution Function of sensitivities (one circle per
branch); inset shows data for the 5\% of branches with the largest
sensitivities. The three lines with the highest sensitivities are
marked with an asterisk. }
\end{figure}

Fig. \ref{fig:Risk-sensitivity} shows the sensitivity of risk to
individual branch outage probabilities (derived from their failure
rates) in the Polish system, as computed from (\ref{eq:sensitivity}). 

Clearly, the sensitivity data in Fig.~\ref{fig:Risk-sensitivity}
show a very heavy tail; a very small number of components affect risk
disproportionately. This becomes particularly clear when we look at
the impact of particular lines on total risk, by computing the risk
sensitivity value times the outage probability. For example, given
the risk sensitivity of $1.41\times10^{5}$ kW and the outage probability
of $1.06\times10^{-4}$ for branch 96 (which has the highest sensitivity
value), and total risk of 50.5 kW in the Polish system (Fig. \ref{fig:mc_and_rc}),
we find that this branch can adjust $29\%$ of overall cascading failure
risk by modifying its failure probability. 

These sensitivity values can be used to identify potential risk-mitigation
strategies. For example, transmission lines that have a higher overall
risk sensitivity factor might be targeted for increased vegetation
management or improved fault detection relaying systems, which may
reduce the associated line-outage probabilities. To illustrate this
approach, we identified the three branches with the highest sensitivity
values, namely branches 96, 23 and 169. After identifying the branches,
we reduced the three failure probabilities ($p_{96}$, $p_{23}$ and
$p_{169}$) by half, and re-computed risk. This resulted in a 33\%
reduction in total cascading failure risk.

\subsection{Estimating risk as a function of load \label{sub:Estimating-risk-vs-load-level}}

Finally, in order to illustrate the potential of this approach to
perform more in-depth studies of cascading failure risk, we used the
Random Chemistry method to examine how cascading failure risk changes
with different load levels and generator dispatch methods. To do so,
we prepared $n-1$ secure versions of both systems for a range of
load levels from 50\% to 119\% for RTS-96 and 50\% to 115\% for the
Polish system. 119\% was the highest load level at which SCDCOPF could
find a solution without load shedding in the pre-contingency RTS-96
case. For the Polish system, load could increase up to 110\% before
a small amount of load shedding was needed to find a secure solution.
For cases from 111\% to 115\% load, less than 1\% load shedding was
required to achieve $n-1$ security. For comparison purposes, we extended
our study of the Polish system up to 115\% load. 

\begin{figure*}
\begin{centering}
\includegraphics[width=1\textwidth]{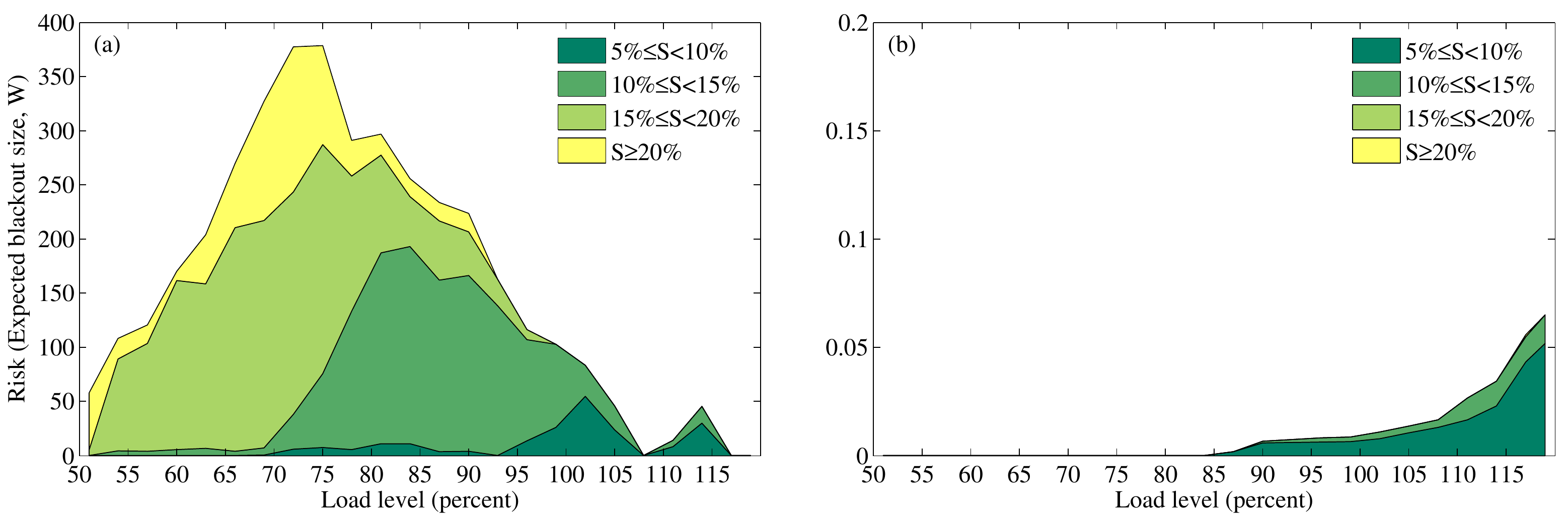}
\par\end{centering}

\protect\caption{\label{fig:Risk-RTS-vs-load-prc}Cascading failure risk vs. load level
for RTS-96: (a) SCDCOPF, and (b) Proportional dispatch.}
\end{figure*}

Fig.~\ref{fig:Risk-RTS-vs-load-prc} shows risk in the RTS-96 system
for all load levels for two different pre-contingency dispatch conditions.
To make the analysis more useful, the risk associated with different
blackout sizes are separately presented. Panel (a) shows the results
produced from a SCDCOPF dispatch at each load level. In order to smooth
out differences at adjacent points, each point on the graph shows
the rolling average of risk across three consecutive integer percentage
load levels (i.e., the datum at 90\% load is the average for 89\%,
90\%, and 91\%). 

It is interesting to note that very-large blackout risk is greatest
at about 70\% load level, and decreases significantly as the load
increases beyond this level. The initial increase in risk is similar
to the findings in \cite{Liao:2004}, \cite{Carreras:2002}, \cite{Chen:2005}
and \cite{Nedic:2006}, which report a phase transition in cascading
failure risk as load increases. However, depending on the dispatch
method for the pre-contingency system, the risk may decrease subsequently,
as Fig.~\ref{fig:Risk-RTS-vs-load-prc}a shows for SCDCOPF dispatch.
The later decrease in risk differs substantially from what is previously
reported in the literature. 

Fig.~\ref{fig:Risk-RTS-vs-load-prc}b shows risk for the same load
levels but with a proportional dispatch method, in which we took the
119\% load case from SCDCOPF, and uniformly decreased the loads and
generators to each lower load level. This dispatch method reduced
risk substantially, and the relationship between load and risk now
becomes monotonic. Large blackout risk is practically zero for the
proportional dispatch cases. The pre-contingency dispatch in this
case is obviously more expensive than that from the SCDCOPF, which
indicates that there is an important tradeoff between generation dispatch
costs and cascading failure risk. 

\begin{figure}
\begin{centering}
\includegraphics[width=1\columnwidth]{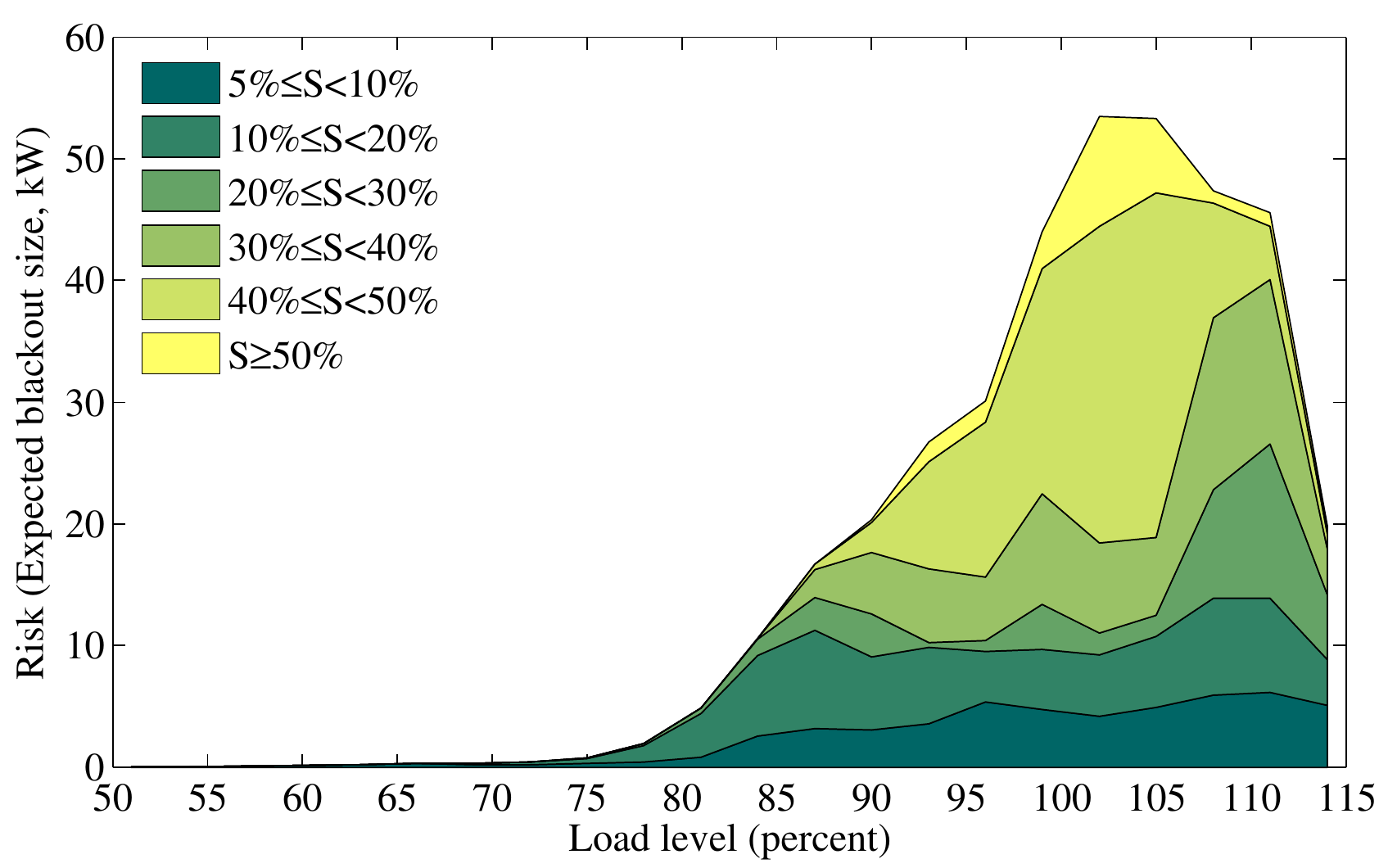}
\par\end{centering}

\protect\caption{\label{fig:Risk-polish-vs-load-prc}Cascading failure risk vs. load
level for the Polish system, dispatched by SCDCOPF.}
\end{figure}

Fig.~\ref{fig:Risk-polish-vs-load-prc} shows cascading failure risk
vs.~load level for the Polish system, using SCDCOPF for the pre-contingency
dispatch. Again, the results suggest that that risk does not always
increase with load level. In fact, risk decreases to some extent for
load percentages between 100\% and 110\%, and to a larger extent above
110\% (the same cases for which some load shedding, less than 1\%,
occurs during the pre-contingency dispatch).

\begin{figure}
\begin{centering}
\includegraphics[width=1\columnwidth]{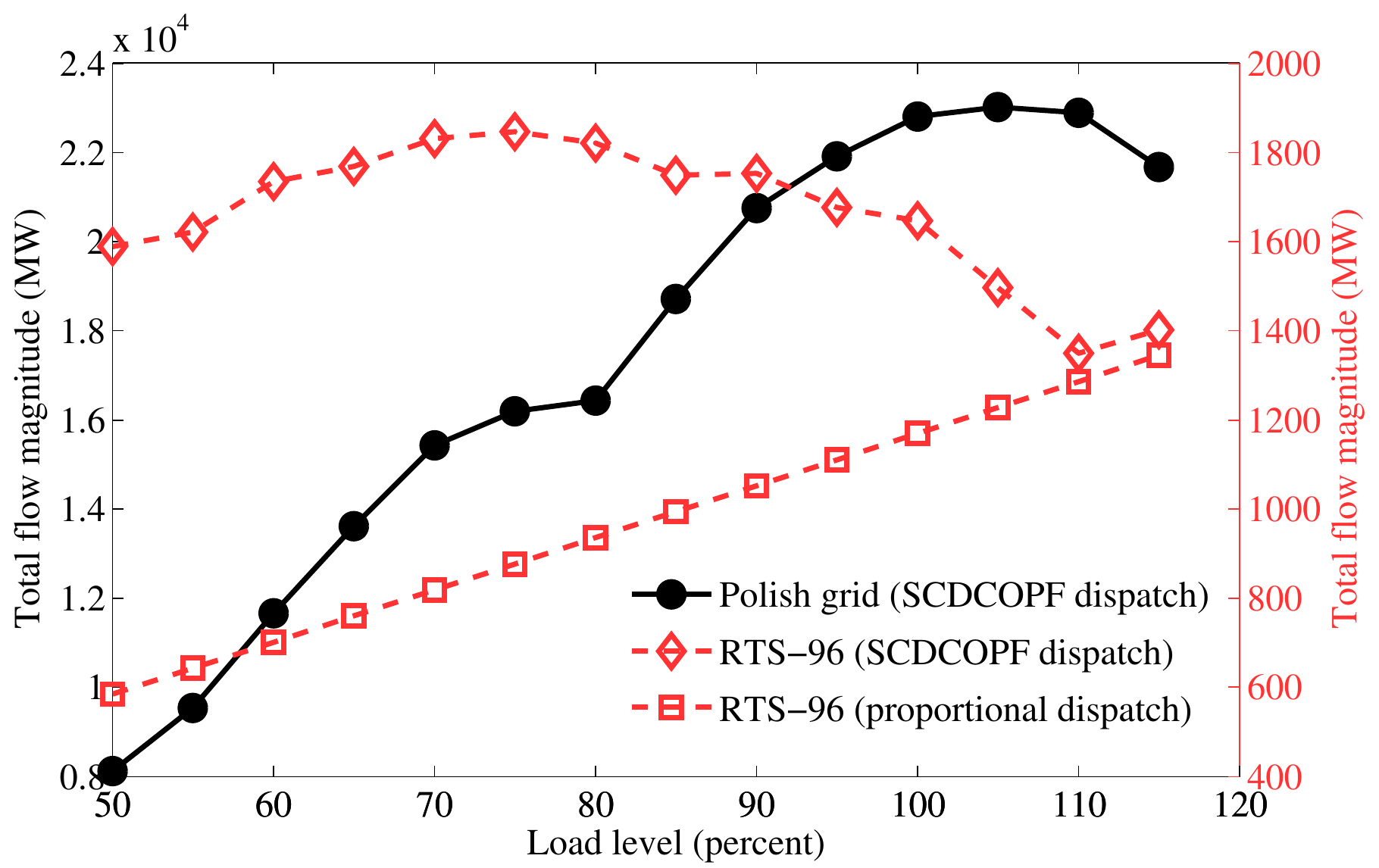}
\par\end{centering}

\protect\caption{\label{fig:sum-flow-vs-load-prc}Total power flow magnitude on the
5\% most sensitive branches in the Polish system (solid circles, left
axis) and RTS-96 (open symbols, right axis). }
\end{figure}

Inspection of the test systems indicates that the reduced risk at
higher load levels results from the way that SCDCOPF uses more local
generation with less long-distance transmission at higher load levels.
In other words, at moderate load levels, the minimum cost generators
are far from load centers, and important transmission corridors are
loaded closer to their capacity. To illustrate this, Fig. \ref{fig:sum-flow-vs-load-prc}
shows the total amount of power flow on the 5\% most sensitive branches,
found from (\ref{eq:sensitivity}), in RTS-96 and the Polish system.
We observe in Fig. \ref{fig:sum-flow-vs-load-prc} that the total
power flows on these critical lines follow the same general pattern
as risk in Figs. \ref{fig:Risk-RTS-vs-load-prc} and \ref{fig:Risk-polish-vs-load-prc}
for both systems. 

\section{Conclusion and Future Work\label{sec:Conclusions}}

This paper presents a new computationally efficient method, based
on the Random Chemistry algorithm in \cite{Eppstein:2012}, for estimating
the risk of large (e.g., $\ge5\%$ of system load) cascading blackouts.
A comparison of this method to Monte~Carlo simulation for two test
cases (the IEEE RTS-96, and a model of the Polish transmission system)
shows that the new approach is at least two orders of magnitude faster
than Monte~Carlo, and does not introduce measurable bias into the
estimate. The computational efficiency of the Random Chemistry approach
comes from the way that it directly searches for large blackout-causing
contingencies, as opposed to Monte~Carlo, which samples broadly from
all possible contingencies. 

We disaggregated risk with respect to the resulting blackout sizes
and the sensitivity of risk to individual branch outage probabilities
in the initiating events. For the latter case, we derived a method
to use the data generated by the Random Chemistry approach to quickly
estimate the impact of component outage probabilities on the overall
risk. For the Polish system, the results indicate that reducing the
unplanned outage probabilities on three transmission lines (e.g.,
by more aggressive vegetation management) would produce a 33\% reduction
in overall cascading failure risk.

In order to illustrate the utility of the proposed approach, we computed
how the risk of blackouts of various sizes changes with load level.
Surprisingly, the results indicate that risk can sometimes decrease
with increasing load, e.g., if generators are dispatched according
to a preventative security constrained optimal power flow. However,
if generators are dispatched proportionally (all generations and loads
change in equal ratios with respect to a load level with low risk),
blackout risk is much smaller and risk increases monotonically with
load. This illustrates an important tradeoff between economic efficiency
and blackout risk.

These results suggest a number of important topics for future research.
First, the results presented here rely on the assumption that branch
outages are uncorrelated and come from a relatively simple cascading
failure simulator. Future work will seek to confirm that the efficiency
gains associated with the Random Chemistry approach persist after
modeling correlated outages and additional mechanisms of cascading.
In addition, we studied only the sensitivity of risk to initiating
events; studying the sensitivity of risk to events during a cascade
remains for future work. Finally, future research is needed to transform
the data that result from Random Chemistry into effective strategies
for reducing blackout risk.

\appendix{}

\section*{I.~Summary of the Random Chemistry Algorithm\label{sec:Overview-of-Random}}

The Random Chemistry algorithm is a stochastic set-size reduction
search strategy that can be used to efficiently (in logarithmic time)
find minimal subsets that are associated with a certain outcome of
interest. The algorithm was applied to genomic data mining in \cite{Eppstein:2007},
and was adapted to the problem of finding $n-k$ blackout-initiating
contingencies in power grids \cite{Eppstein:2012}. In summary, the
Random Chemistry algorithm operates as follows. Initially, we use
the cascading failure simulator from \cite{Eppstein:2012} (DCSIMSEP
\cite{DCSIMSEP}) to find a large multiple contingency $c$ (an $n-k_{\textrm{init}}$
contingency, where $k_{\textrm{init}}\ge40$) that results in a large
blackout. Because $k_{\textrm{init}}$ is so large, this step typically
requires very few tries. Then, the algorithm stochastically reduces
$c$ according to a logarithmically decreasing set size reduction
schedule (in this work we used $\{$$k_{\textrm{init}}=80$, $k_{2}=40$,
$k_{3}=20$, $k_{4}=14$, $k_{5}=10$, $k_{6}=7$, $k_{\textrm{final}}=5$$\}$)
by testing random subsets of the desired size until one is found that
causes a large blackout. If no such subset is found within a pre-specified
maximum number of tries $T$ (we used $T=20$), the run is restarted
from a new random $n-k_{\textrm{init}}$ contingency. The remaining
set of size $k_{\textrm{final}}=5$ is exhaustively searched (starting
from $k=2$ ) until a minimal $n-k$ blackout-causing contingency
($2\le k\le k_{\textrm{final}}$) is identified. This cycle can then
be repeated to obtain large collections of $n-k$ minimal hazardous
contingencies.

\section*{II.~Exploring Assumption 1}

\begin{figure}
\begin{centering}
\includegraphics[width=1\columnwidth]{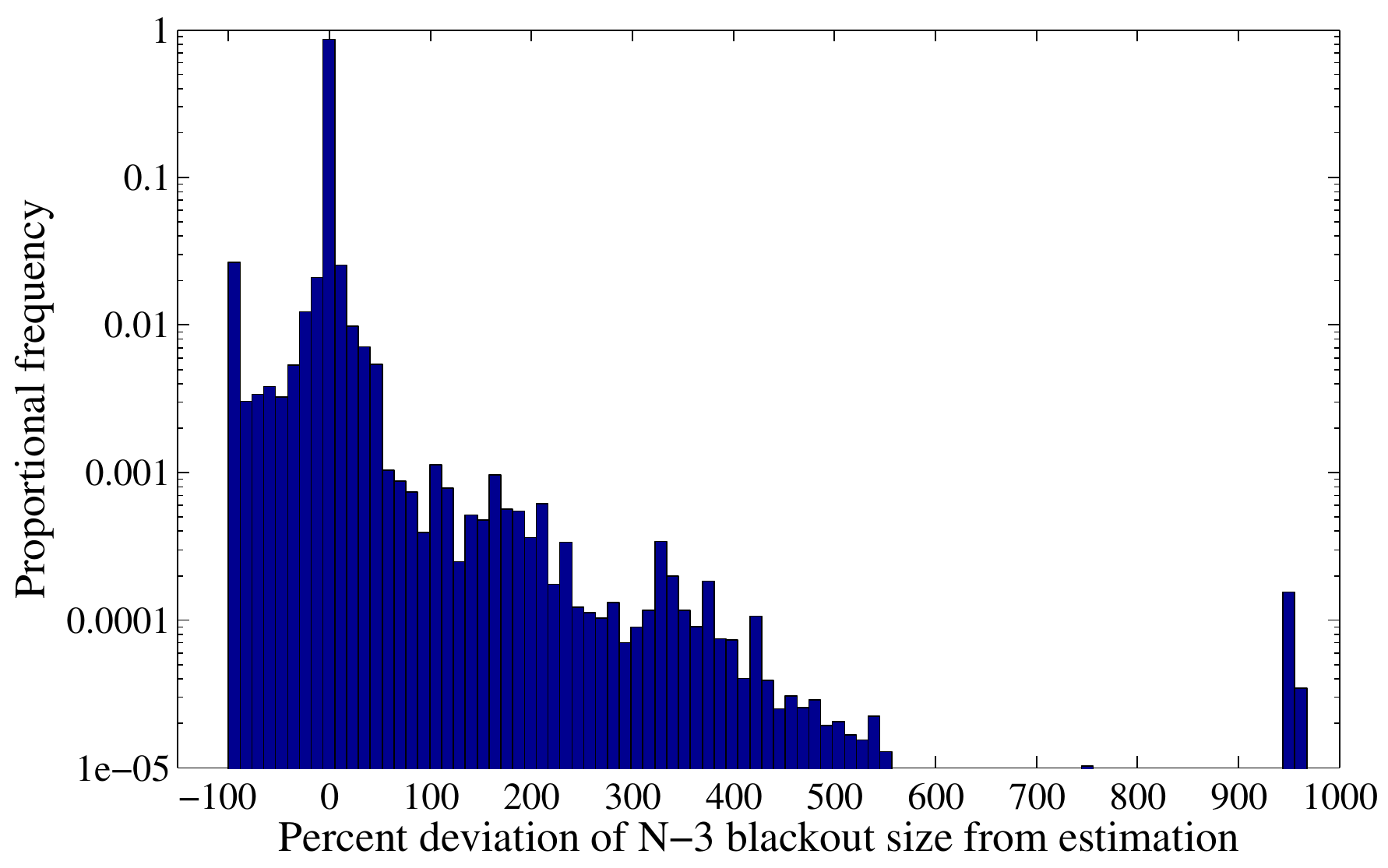}
\par\end{centering}

\protect\caption{\label{fig:assumption1}Proportional frequency of percent difference
in actual minus estimated blackout sizes caused by non-minimal $n-3$
contingencies. }
\end{figure}

In this section, we test Assumption 1 for the Polish system by comparing
the actual blackout sizes of all $n-3$ supersets of all $n-2$ malignancies
with the estimated values, i.e., the blackout size of each $n-2$
subset. For each $n-2$ malignancy, there are $n-2=2894$ supersets
that were simulated with DCSIMSEP for this comparison. We then computed
the percent deviation of each $n-3$ blackout size from its malignancy
subset:

\[
\epsilon=\frac{S_{act}-S_{est}}{S_{est}}\times100
\]
where $S_{act}$ is the actual blackout size of an $n-3$ superset,
and $S_{est}$ is the blackout size of its $n-2$ subset. Fig. \ref{fig:assumption1}
shows the proportional frequency of these deviations. We found that
these deviations are generally in the range of \textcolor{black}{-1.9\%
($10^{th}$ percentile) to 0.9\% ($90^{th}$ percentile). Ultimately,
Assumption 1 resulted in the estimated $n-3$ blackout risk to be
2.5\% higher than the exact value obtained by using the actual $n-3$
blackout sizes.}

Note that Assumption 1 provides a weaker estimation when larger than
one-extra-element supersets of malignancies are considered. \textcolor{black}{However,
with larger supersets, the contingencies become highly improbable,
and make a negligible contribution to the total risk. Thus, we only
need Assumption 1 to work for supersets with few additional elements
added to a malignancy. This partially explains why the above assumption
does not }measurably change the Random Chemistry risk estimation results
in section \ref{sec:Results}, relative to what is computed by Monte~Carlo.

\section*{Acknowledgment}

The authors gratefully acknowledge helpful conversations with Ian
Dobson about these results, and Mark Wagy for assistance with an earlier
version of this method.

\bibliographystyle{ieeetr}
\bibliography{RC_risk}

\begin{thebibliography}{10}

\bibitem{:2004}
Final report on the {A}ugust 14, 2003 blackout in the {U}nited {S}tates and
  {C}anada.
\newblock Technical report, US-Canada Power System Outage Task Force, 2004.

\bibitem{:2012}
Arizona-{S}outhern {C}alifornia outages on {S}eptember 8, 2011.
\newblock Technical report, FERC and NERC, 2012.

\bibitem{NERC:2007}
NERC Standard TOP-004-2.
\newblock {\em Transmission Operations}, 2007.

\bibitem{Vaiman:2012}
M.~{M. Vaiman, \emph{et al.}}
\newblock Risk assessment of cascading outages: Methodologies and challenges.
\newblock {\em IEEE Transactions on Power Systems}, 27(2):631--641, 2012.

\bibitem{Allan:1996}
R.~N. Allan and R.~Billinton.
\newblock {\em Reliability Evaluation of Power Systems}.
\newblock Plenum Press, 1996.

\bibitem{Borges:2001}
{C.L.T. Borges \emph{et al.}}
\newblock Composite reliability evaluation by sequential {M}onte {C}arlo
  simulation on parallel and distributed processing environments.
\newblock {\em IEEE Transactions on Power Systems}, 16(2):203--209, 2001.

\bibitem{Billinton:1997}
R.~Billinton and A.~Jonnavithula.
\newblock Composite system adequacy assessment using sequential {M}onte {C}arlo
  simulation with variance reduction techniques.
\newblock {\em IEE Proceedings-Generation, Transmission and Distribution},
  144(1):1--6, 1997.

\bibitem{Singh:1997}
C.~Singh and J.~Mitra.
\newblock Composite system reliability evaluation using state space pruning.
\newblock {\em IEEE Transactions on Power Systems}, 12(1):471--479, 1997.

\bibitem{DaSilva:2000}
{A.M. Leite da Silva, \emph{et al.}}
\newblock Pseudo-chronological simulation for composite reliability analysis
  with time varying loads.
\newblock {\em IEEE Transactions on Power Systems}, 15(1):73--80, 2000.

\bibitem{daSilva:2007}
{A.M. Leite da Silva, \emph{et al.}}
\newblock Composite reliability assessment based on {M}onte {C}arlo simulation
  and {A}rtificial {N}eural {N}etworks.
\newblock {\em IEEE Transactions on Power Systems}, 22(3):1202--1209, 2007.

\bibitem{Henneaux:2013}
P.~Henneaux, P.-E. Labeau, and J.-C. Maun.
\newblock Blackout probabilistic risk assessment and thermal effects: Impacts
  of changes in generation.
\newblock {\em IEEE Transactions on Power Systems}, 28(4):4722--4731, Nov 2013.

\bibitem{Carreras:2000}
{B. A. Carreras \emph{et al.}}
\newblock Initial evidence for self-organized criticality in electric power
  system blackouts.
\newblock In {\em Proceedings of Hawaii International Conference on System
  Sciences}, 2000.

\bibitem{Carreras:2002}
{B. A. Carreras \emph{et al.}}
\newblock Critical points and transitions in an electric power transmission
  model for cascading failure blackouts.
\newblock {\em Chaos: An interdisciplinary journal of non-linear science},
  12(4):985--994, 2002.

\bibitem{Dobson:2013}
I.~Dobson, B.A. Carreras, and D.E. Newman.
\newblock How many occurrences of rare blackout events are needed to estimate
  event probability?
\newblock {\em IEEE Transactions on Power Systems}, 28(3):3509--3510, {Aug}
  2013.

\bibitem{Pfitzner:2011}
R.~Pfitzner, Konstantin Turitsyn, and Michael Chertkov.
\newblock Statistical classification of cascading failures in power grids.
\newblock In {\em IEEE Power and Energy Society General Meeting}, pages 1--8,
  July 2011.

\bibitem{Qi:2013}
Junjian Qi, Shengwei Mei, and Feng Liu.
\newblock Blackout model considering slow process.
\newblock {\em IEEE Transactions on Power Systems}, 28(3):3274--3282, Aug 2013.

\bibitem{Rahnamay-Naeini:2014}
{M. Rahnamay-Naeini, \emph{et al.}}
\newblock Stochastic analysis of cascading-failure dynamics in power grids,
  2014.

\bibitem{Mousavi:2012}
O.~Alizadeh Mousavi, R.~Cherkaoui, and M.~Bozorg.
\newblock Blackouts risk evaluation by {M}onte {C}arlo simulation regarding
  cascading outages and system frequency deviation.
\newblock {\em Electric Power Systems Research}, 89(0):157 -- 164, 2012.

\bibitem{Kirschen:2004}
{D.S. Kirschen \emph{et al.}}
\newblock A probabilistic indicator of system stress.
\newblock {\em IEEE Transactions on Power Systems}, 19(3):1650--1657, 2004.

\bibitem{Chen:2005}
J.~Chen, J.~S. Thorp, and I.~Dobson.
\newblock {Cascading dynamics and mitigation assessment in power system
  disturbances via a hidden failure model}.
\newblock {\em International Journal of Electrical Power \& Energy Systems},
  27(4):318--326, 2005.

\bibitem{Kim:2013}
J.~Kim, J.A. Bucklew, and I.~Dobson.
\newblock Splitting method for speedy simulation of cascading blackouts.
\newblock {\em IEEE Transactions on Power Systems}, 28(3):3010--3017, 2013.

\bibitem{Chen:2013}
Q.~Chen and L.~Mili.
\newblock Composite power system vulnerability evaluation to cascading failures
  using importance sampling and antithetic variates.
\newblock {\em IEEE Transactions on Power Systems}, 28(3):2321--2330, 2013.

\bibitem{Ren:2008}
Hui Ren and Ian Dobson.
\newblock Using transmission line outage data to estimate cascading failure
  propagation in an electric power system.
\newblock {\em IEEE Transactions on Circuits and Systems--II: Express Briefs},
  55(9):927--931, 2008.

\bibitem{Dobson:2012}
I.~Dobson.
\newblock Estimating the propagation and extent of cascading line outages from
  utility data with a branching process.
\newblock {\em IEEE Transactions on Power Systems}, 27(4):2146--2155, 2012.

\bibitem{Eppstein:2012}
M.J. Eppstein and P.D.H. Hines.
\newblock A "{R}andom {C}hemistry" algorithm for identifying collections of
  multiple contingencies that initiate cascading failure.
\newblock {\em IEEE Transactions on Power Systems}, 27(3):1698--1705, 2012.

\bibitem{Rezaei:2014}
P.~Rezaei, P.D.H. Hines, and M.J. Eppstein.
\newblock Estimating cascading failure risk: Comparing {M}onte {C}arlo sampling
  and random chemistry.
\newblock In {\em IEEE Power and Energy Society General Meeting}, pages 1--5,
  July 2014.

\bibitem{Rezaei:2014aa}
P.~Rezaei and P.D.H. Hines.
\newblock Changes in cascading failure risk with generator dispatch method and
  system load level.
\newblock In {\em IEEE PES Transmission and Distribution Conference and
  Exposition}, pages 1--5, 2014.

\bibitem{Cochran:1963}
W.~G. Cochran.
\newblock {\em Sampling Techniques}.
\newblock Wiley, 2nd edition, 1963.

\bibitem{Grigg:1999}
{C. Grigg, \emph{et al.}}
\newblock The {IEEE} reliability test system-1996. a report prepared by the
  reliability test system task force of the application of probability methods
  subcommittee.
\newblock {\em IEEE Transactions on Power Systems}, 14(3):1010--1020, 1999.

\bibitem{Nedic:2006}
{D.P. Nedic, \emph{et al.}}
\newblock Criticality in a cascading failure blackout model.
\newblock {\em International Journal of Electrical Power and Energy Systems},
  28(9):627 -- 633, 2006.

\bibitem{Zimmerman:2011}
R.D. Zimmerman, C.E. Murillo-S{\'a}nchez, and R.J. Thomas.
\newblock {MATPOWER}: Steady-state operations, planning, and analysis tools for
  power systems research and education.
\newblock {\em IEEE Transactions on Power Systems}, 26(1):12 --19, feb. 2011.

\bibitem{EIA}
http://www.eia.gov/.

\bibitem{DCSIMSEP}
P.D.H. Hines.
\newblock {DCSIMSEP}: A simulator of cascading separation in power grids.
  \url{http://uvm.edu/~phines/dcsimsep/}.

\bibitem{Liao:2004}
H.~Liao, J.~Apt, and S.~Talukdar.
\newblock Phase transitions in the probability of cascading failures.
\newblock In {\em Electricity Transmission in Deregulated Markets: Conference
  at Carnegie Mellon University}, Pittsburgh, PA, 2004.

\bibitem{Eppstein:2007}
{M.J. Eppstein \emph{et al.}}
\newblock Genomic mining for complex disease traits with ``{R}andom
  {C}hemistry''.
\newblock {\em Genetic Programming and Evolvable Machines}, 8(4):395--411,
  2007.

\end{thebibliography}

\section*{Author Biographies}

\begin{IEEEbiography}[{\includegraphics[width=1in,height=1.25in,clip,keepaspectratio]{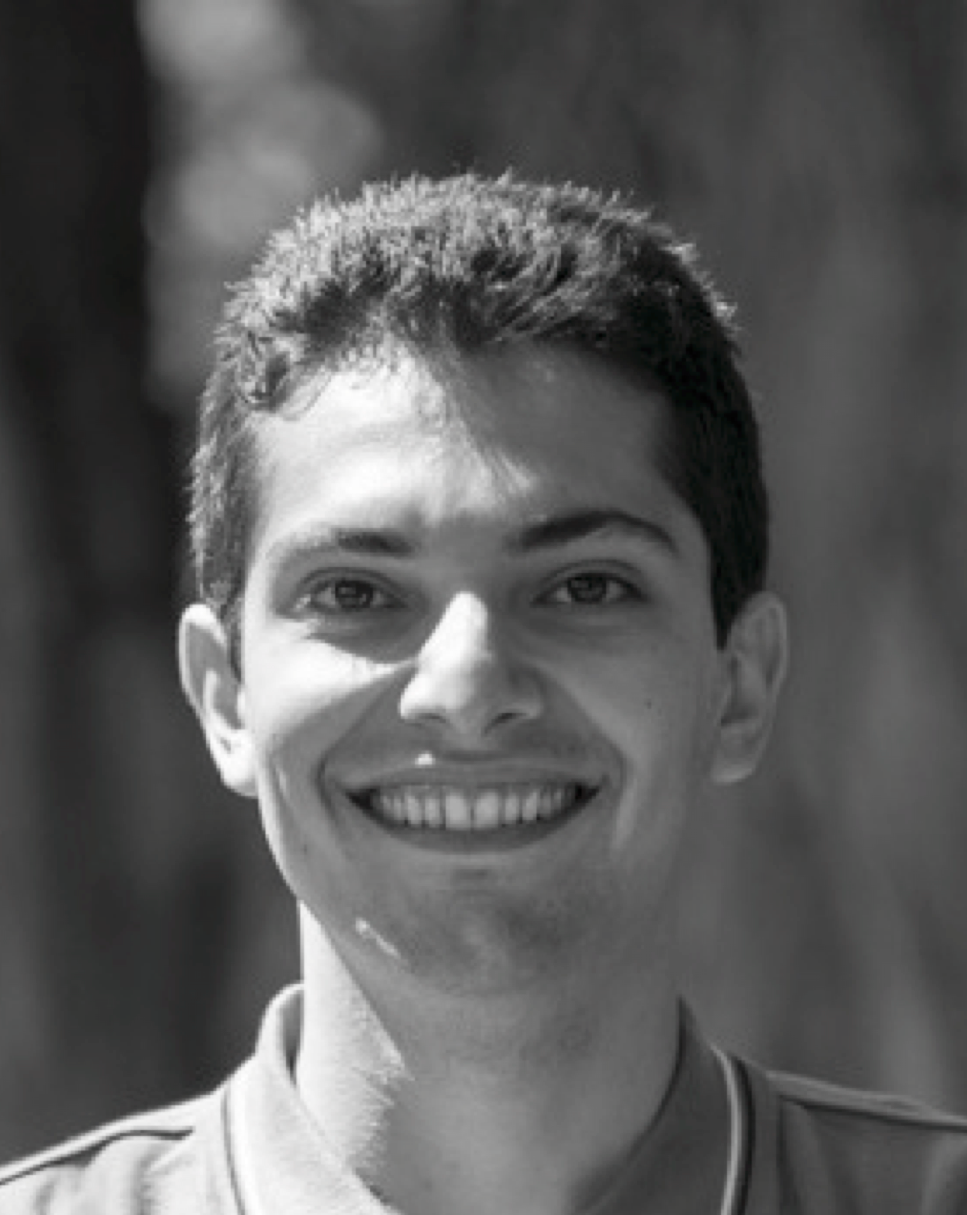}}]{Pooya Rezaei} (S'12) received the M.Sc.~and B.Sc.~degrees in Electrical Engineering from Sharif University of Technology (2010) and University of Tehran (2008) respectively. Currently, he is pursuing the Ph.D. in Electrical Engineering at the University of Vermont.  \end{IEEEbiography}

\begin{IEEEbiography}[{\includegraphics[width=1in,height=1.25in,clip,keepaspectratio]{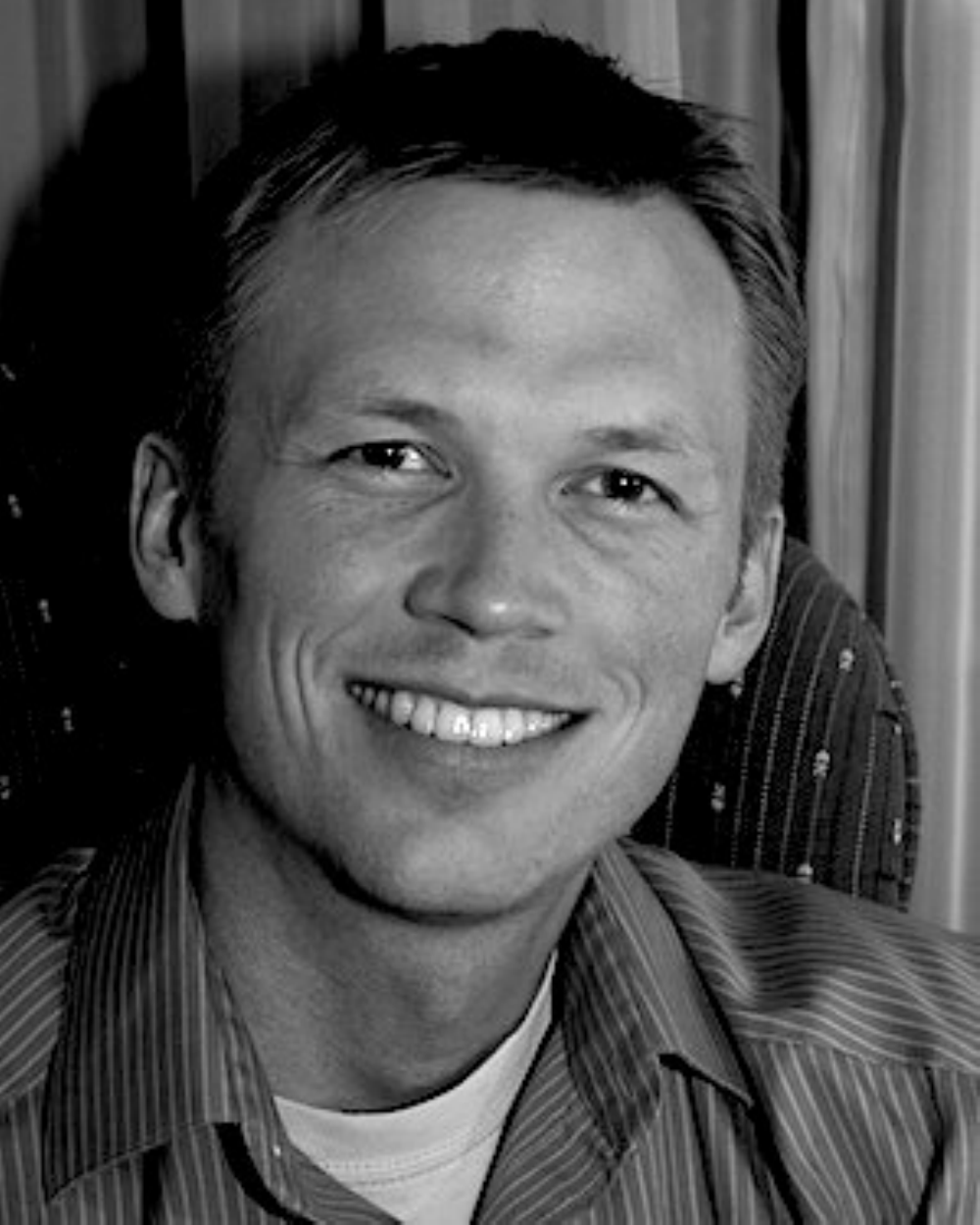}}]{Paul D. H. Hines} (S'96,M'07,SM'14) received the Ph.D. in Engineering and Public Policy from Carnegie Mellon University in 2007 and M.S. (2001) and B.S. (1997) degrees in Electrical Engineering from the University of Washington and Seattle Pacific University, respectively. He is currently an Associate Professor in the School of Engineering at the University of Vermont.
\end{IEEEbiography}

\begin{IEEEbiography}[{\includegraphics[width=1in,height=1.25in,clip,keepaspectratio]{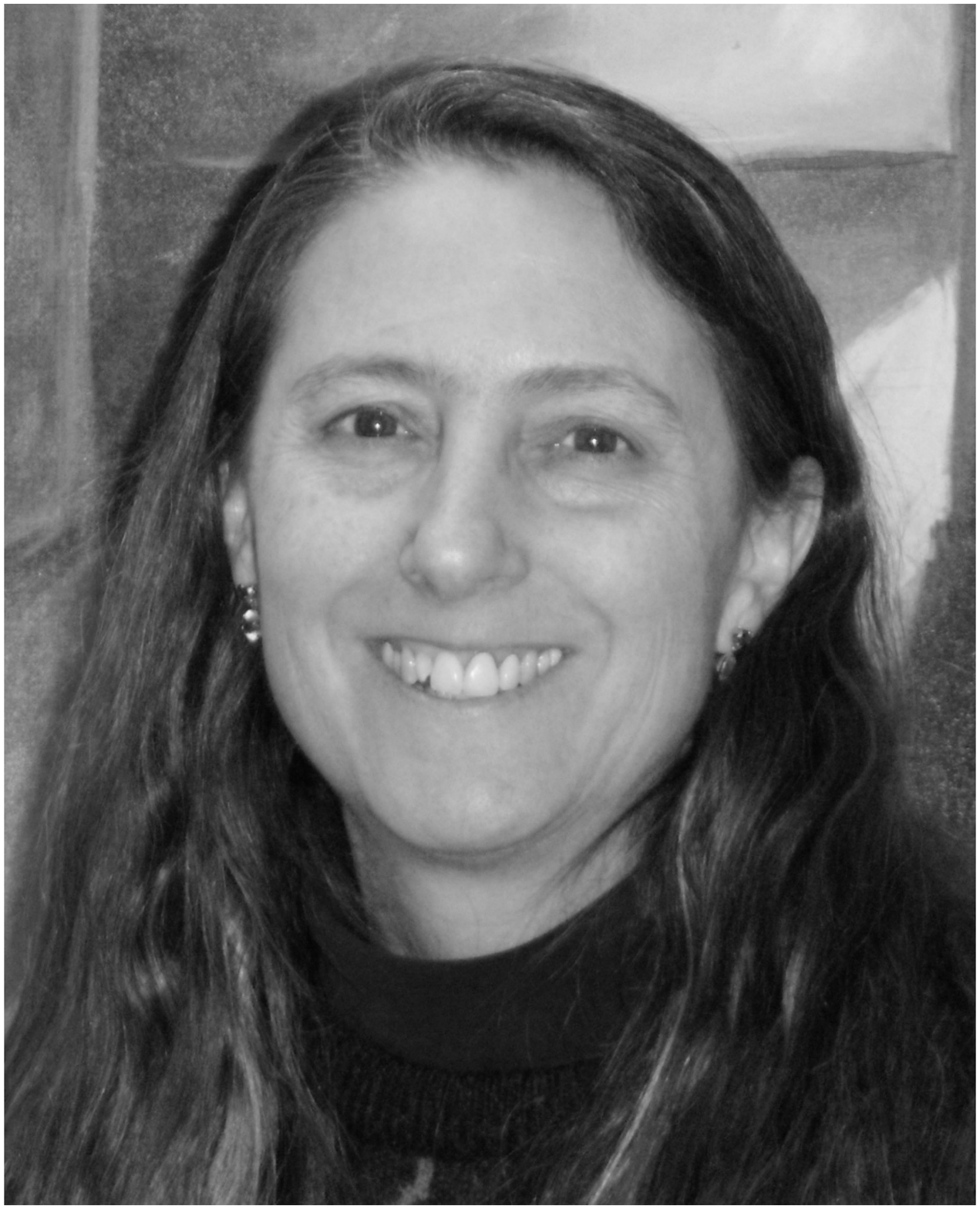}}]{Margaret J. Eppstein} is Professor and Chair of Computer Science at the University of Vermont (UVM), where she has been on the faculty since 1983. She received a B.S. in Zoology from Michigan State University in 1978, an M.S. in Computer Science from UVM in 1983, and a Ph.D. in Environmental Engineering at UVM in 1997. She was the founding director of the Vermont Complex Systems Center (2006-2010).
\end{IEEEbiography}

\vfill
\end{document}